\documentclass[conference, keeplastbox]{IEEEtran}
\IEEEoverridecommandlockouts

\usepackage{cite}
\usepackage{amsmath,amssymb,amsfonts, mathtools, array}

\usepackage{graphicx}
\usepackage{textcomp}
\usepackage{xcolor}
\usepackage{algorithm}
\usepackage{algpseudocode}
\usepackage{caption}
\usepackage{subcaption}
\usepackage{tabularx}
\usepackage{float}
\usepackage{lipsum,mathtools}
\usepackage{multirow}
\usepackage{dirtytalk}
\usepackage[numbers]{natbib}

\def\BibTeX{{\rm B\kern-.05em{\sc i\kern-.025em b}\kern-.08em
    T\kern-.1667em\lower.7ex\hbox{E}\kern-.125emX}}
    \usepackage[left=0.57in,right=0.58in,top=0.7in,bottom=0.95in]{geometry}
\begin{document}

\title{Enhancing IoT Security: A Novel Feature Engineering Approach for ML-Based Intrusion Detection Systems \\
}
\author{\IEEEauthorblockN{Afsaneh Mahanipour}
\IEEEauthorblockA{\textit{Department of Computer Science} \\
\textit{University of Kentucky}\\
Lexington, KY, USA \\
ama654@uky.edu}
\and
\IEEEauthorblockN{Hana Khamfroush}
\IEEEauthorblockA{\textit{Department of Computer Science} \\
\textit{University of Kentucky}\\
Lexington, KY, USA \\
khamfroush@cs.uky.edu}
}
\maketitle
\begin{abstract}
The integration of Internet of Things (IoT) applications in our daily lives has led to a surge in data traffic, posing significant security challenges. IoT applications using cloud and edge computing are at higher risk of cyberattacks because of the expanded attack surface from distributed edge and cloud services, the vulnerability of IoT devices, and challenges in managing security across interconnected systems leading to oversights. This led to the rise of ML-based solutions for intrusion detection systems (IDSs), which have proven effective in enhancing network security and defending against diverse threats. However, ML-based IDS in IoT systems encounters challenges, particularly from noisy, redundant, and irrelevant features in varied IoT datasets, potentially impacting its performance. Therefore, reducing such features becomes crucial to enhance system performance and minimize computational costs. This paper focuses on improving the effectiveness of ML-based IDS at the edge level by introducing a novel method to find a balanced trade-off between cost and accuracy through the creation of informative features in a two-tier edge-user IoT environment. A hybrid Binary Quantum-inspired Artificial Bee Colony and Genetic Programming algorithm is utilized for this purpose. Three IoT intrusion detection datasets, namely NSL-KDD, UNSW-NB15, and BoT-IoT, are used for the evaluation of the proposed approach. Performance analysis is conducted using various evaluation metrics such as accuracy, sensitivity, specificity, and False Positive Rate (FPR) are employed, while the cost of the IDS system is assessed based on computational time. The results are compared with existing methods in the literature, revealing that the IDS performance can be enhanced with fewer features, consequently reducing computational time, through the proposed method. This offers a better performance-cost trade-off for the IDS system.
\end{abstract}


\section{Introduction}
In the era of Internet of Things (IoT) applications and widespread internet usage, security has become a major and growing concern. The interconnected nature of IoT devices has expanded the attack surface, providing numerous entry points for malicious actors. The increasing volume of data exchanged between these devices poses a significant target for cybercriminals, raising the potential impact of security breaches across various aspects of daily life, from smart homes to industrial systems. By the end of 2024, it is anticipated that there will be 83 billion IoT devices, utilized across various domains such as intelligent transportation, smart healthcare, and others, contributing to the development of smart cities \cite{albulayhi2022iot}. 

This surge in IoT adoption has also led to challenges in data management and processing. The data collected by IoT devices has been transmitted to a distant cloud server for further processing. However, this setup poses challenges due to the significant distance between the devices and the server, resulting in delays that are not conducive to the time-sensitive nature of IoT applications. Therefore, with the introduction of an edge computing framework, data is sent to edge servers located closer to the IoT devices in most applications. This reduces latency, but it's important to note that these edge servers still face limitations in terms of storage and computational power compared to cloud servers \cite{mahanipour2023wrapper}.

Moreover, within IoT frameworks, numerous devices interact through web application interfaces, necessitating robust authentication and encryption methods to ensure secure communication. In this context, intrusion detection systems (IDSs) play a crucial role in monitoring and identifying attacks. IDSs are broadly categorized into two types based on their discovery methods: signature-based IDS and anomaly-based IDS. The signature-based method relies on pre-stored rules that represent specific attack types. Consequently, any attack not included in the pre-stored rules will go undetected. Conversely, anomaly-based IDS exhibits proficiency in efficiently detecting zero-day (unknown) attacks \cite{mishra2023mitigating}.

Lately, the application of machine learning (ML) techniques has become increasingly prevalent in the field of intrusion detection within IoT IDSs. However, IoT devices exhibit differences in their hardware attributes, functions, and computational capabilities to generate features. When IoT devices transmit their data to an edge server, some features may be noisy, irrelevant or redundant. These characteristics significantly affect the performance of IDSs \cite{albulayhi2022iot}. Selecting informative features is pivotal in ML-based methods for enhancing IDS accuracy by eliminating non-informative features and reducing complexity costs through dataset size reduction. While some studies, such as \cite{al2022wrapper, kareem2022effective}, have employed feature selection (FS) techniques for this purpose, the exploration of constructing new high-level features for IDSs remains unaddressed in existing literature. This paper proposes a novel method to find a balanced trade-off between costs and accuracy by constructing new informative features for ML-based IDSs within the diverse IoT environment.

The search space for feature construction is extensive, as informative features must be combined with appropriate operators to create distinctive new features. Therefore, selecting prominent features for this purpose can be beneficial. To achieve this, the binary quantum-inspired artificial bee colony (BQABC) algorithm \cite{barani2018bqiabc} is used for selecting informative features. BQABC has better convergence rate and exploration capability to prevent trapping in a local optima compared to other binary optimization algorithms such as binary quantum-inspired particle swarm optimization (BQIPSO) and binary quantum-inspired evolutionary algorithm (BQIEAo). Subsequently, genetic programming (GP) \cite{koza1994genetic}, a domain-independent problem-solving approach, is utilized to construct high-level features for more accurate intrusion detection with less computation cost. The main contributions of the proposed method can be summarized in the following manner:

\begin{itemize}
\item Utilizing feature construction method for ML-based IDSs for the first time
\item Proposing a novel feature engineering method by integrating FS based on BQABC algorithm  for the identification of important features and FC based on GP for constructing new feature to create a new dataset
\item Evaluating the performance of the proposed method in terms of accuracy, sensitivity, specificity, False Positive Rate (FPR), number of features, and computational time
\item Comparing the proposed method with seven other methods in the literature by employing various evaluation metrics and obtaining better results such as improving accuracy by approximately \(8\%\) and reducing computational time by an average of 2948 \(s\) across all three datasets
\end{itemize}

\section{BACKGROUND AND RELATED WORKS}

\subsection{Related Works}
Supervised ML classification algorithms use training data to establish a functional relationship between input features and class values. The trained classifier then predicts class values for query instances. However, the dataset may contain redundant, noisy, or irrelevant features, which can harm classification performance. To address this, feature engineering techniques like feature selection (FS), and feature construction (FC) are employed to enhance feature quality and boost classifier accuracy. FC creates new features through functional expressions to improve performance and reveal hidden relationships, while FS selects informative features and reduces the total number by eliminating non-informative ones \cite{mahanipour2019multiple}.

While FS methods have been used in ML-based IDSs, FC techniques have not been applied. Consequently, this section investigates previous research studies that have employed various FS methods to enhance the performance of ML-based IDSs. FS methods can be mainly categorized into filter and wrapper methods. Filter methods operate independently of the learning algorithms and rank features based on their inherent characteristics like information theory, mutual information, or correlation criteria \cite{mishra2023mitigating}. For example, in \cite{thakkar2023fusion}, the features were ranked by combining their statistical importance through Standard Deviation and the Difference of Mean and Median. In another study \cite{kumar2020integrated}, information gain was used as FS method. 

In contrast, wrapper methods employ learning algorithms to assess the quality of selected features, making them more accurate than filter methods. For instance, in \cite{liu2022hybrid}, Genetic algorithm with Random Forest-based fitness function was used to select informative features. Similarly, in \cite{al2022wrapper}, differential evaluation algorithm was used to select the most suitable features and subsequently assessed the selected features using the extreme learning machine classifier. In the paper by Kareem et al. \cite{kareem2022effective}, they employed Gorilla Troops Optimizer as a FS algorithm and augmented its exploitation capability by integrating the bird swarms algorithm. In another study \cite{nazir2021novel}, the authors employed the Tabu Search algorithm in conjunction with a random forest classifier to identify the optimal subset of features for IDSs.

Reviews indicate the need for more optimal solutions due to low classification accuracy in existing methods. Additionally, FC methods have not been employed as a pre-processing step in IDSs. Therefore, this paper proposes a hybrid feature engineering method to find a balanced trade-off between computational cost and accuracy.

\subsection{Binary Quantum-inspired Artificial Bee Colony}
Binary Quantum-inspired Artificial Bee Colony (BQABC) is proposed by Barani and Nezamabadi-pour in 2017 \cite{barani2018bqiabc}. It combines Artificial Bee Colony (ABC)'s main structure with quantum computing principles, offering high exploration capability and robustness for binary optimization problems. 

In the BQABC, some concepts of quantum computing like quantum bits and quantum gates are applied in the main structure of ABC algorithm to define the position of food sources and their updating process. The pseudocode of the BQABC algorithm is given in Algorithm 1. You can read more details about this algorithm in \cite{barani2018bqiabc}.

\begin{algorithm}
\caption{Pseudocode of the BQABC algorithm}\label{alg:cap}
\begin{algorithmic}[1]
\renewcommand{\algorithmicrequire}{\textbf{Input:}}
\renewcommand{\algorithmicensure}{\textbf{Output:}}
\Require The number of population and Maximum iterations (termination condition) 
\Ensure Best food source
\newline
\State Initialize quantum food sources (\(Q(t)\)) randomly, a set of best food sources \(FB(t)=\{\}\), a set of current food sources \(FW(t)=\{\}\), and \(t=0\)
\While {not termination condition}
    \State Observe \(Q(t)\) and make \(FW(t)\)
    \State Calculate fitness values of \(F_i (t)\in FW(t)\), by a desired fitness function (\(fit(F_i)\))
    \State Update \(FB(t)\)
    \For {each employed bee \(i\)}
        \State Generate a new quantum food source \(q_i^\prime\) in the neighborhood of \(q_i\) using Equations (12,13,14) in \cite{barani2018bqiabc}.
        \State Observe \(q_i^\prime\) and make \(F_i^\prime\)
        \State Calculate fitness value of \(F_i^\prime\)
        \If{$fit(F_i^\prime)>fit(F_i)$}
            \State $q_i=q_i^\prime$
            \State $F_i=F_i^\prime$
    \EndIf
    \EndFor
    \For {each onlooker bee \(j\)}
        \State Calculate the probability of food sources using Eq. 4 in \cite{barani2018bqiabc}.
        \State Select a quantum food source \(q_j\) based on probability values
        \State Generate a new quantum food source \(q_j^\prime\)  in the neighborhood of \(q_j\) using Equations (12,13,14) in \cite{barani2018bqiabc}.
        \State Observe \(q_j^\prime\) and make \(F_j^\prime\)
        \State 	Calculate fitness value of \(F_j^\prime\)
        \If{$fit(F_j^\prime)>fit(F_j)$}
            \State $q_j=q_j^\prime$
            \State $F_j=F_j^\prime$
    \EndIf
    \EndFor
    \State Determine abandoned food source and replace it with a new quantum food source for the scout bee
    \State Memorize the best food source found so far
    \State $t=t+1$
\EndWhile
\end{algorithmic}
\end{algorithm}

\subsection{Genetic Programming (GP)}
Genetic Programming (GP) was introduced for the first time by John Koza \cite{koza1994genetic}. This algorithm used genetic algorithm as a process to evolve mathematical functions, but chromosomes are encoded with tree structure. The leaves of trees can be selected from the variables of the problem, and internal nodes can be selected from predefined mathematical operators. GP is a population-based evolutionary algorithm that tries to find an optimum function for the desired problem. For this purpose, it generates a number of chromosomes with different sizes, and then selects the best one based on their fitness values. The GP search procedure can be narrated by several main steps:

\begin{enumerate}
  \item Initializing a random population of individual chromosomes using variables and operators.
  \item Iterating the following sub-steps until reaching a stopping criterion:
  \begin{enumerate}
    \item Evaluation: calculating fitness value of each chromosome by an appropriate fitness function.
    \item Selection: choosing one or more chromosomes of the population by a selection approach to participate in the next sub-step.
    \item Evolution: generating new chromosomes and developing a new population by applying genetic operators including: reproduction, crossover, and mutation on the selected chromosomes.
  \end{enumerate}
  \item Returning the best chromosome with the maximum fitness value as the optimum solution.
\end{enumerate}

\section{Proposed Method}

In this section, the proposed feature engineering method for intrusion detection is explained. This method utilizes both FS and FC techniques to provide informative features for precise classification of network attacks, while also finding the trade-off between accuracy and computational cost. To achieve this, the BQABC nature-inspired algorithm is employed to evaluate dataset features and eliminate non-informative ones, thereby reducing the search space and improving the quality of the feature set. For example, consider a dataset \(D=\{X, Y\}:=\{(x_n, y_n)\}_{n=1}^N\), consisting of \(N\) samples and a feature set \(F=\{f_1, f_2, ..., f_m\}\). Here, \(x_n=(x_{n1}, x_{n2}, ..., x_{nm})\) represents a sample vector, and \(Y=(y_{1}, y_{2}, ..., y_{N})\) denotes the class label vector of \(N\) samples, where \(y_n\in \{1, ..., ClassLabel\}\). \(F=\{f_1, f_2, ..., f_m\}\) refers the original feature set used by BQABC to select informative features \(Selected-F=\{f_1, f_2, ..., f_s\}\). Subsequently, the selected features are inputted into the GP algorithm, a type of FC algorithm, to generate a new, higher-level, and more distinctive feature \(\{f_{c}\}\). This newly constructed feature is then added to the selected feature subset, resulting in an augmented feature set \(Augmented-F=\{f_1, f_2, ..., f_s, f_{c}\}\). Finally, this augmented feature set is fed into a Machine Learning (ML) classifier to detect security attacks.

\textbf{System Architecture:} In this paper, intrusion attack detection is performed at the edge level. Data collected by the IoT devices are sent to their closest edge server for further processing including intrusion detection. As previously stated, the distance (\(D\)) between IoT devices and the cloud server is significantly greater than the distance (\(d\)) between IoT devices and the nearest edge server. Therefore, by transmitting data to the edge server, latency is reduced, making it more suitable for time-sensitive IoT applications. However, edge servers have inherent limitations in terms of storage and computational power compared to cloud servers. As illustrated in Fig. 1, our focus is solely on detecting attacks on a single edge server, assuming that each edge server is tasked to perform intrusion detection for its assigned set of IoT devices.

\begin{figure*}[!ht]
\centering
\includegraphics[width=\textwidth]{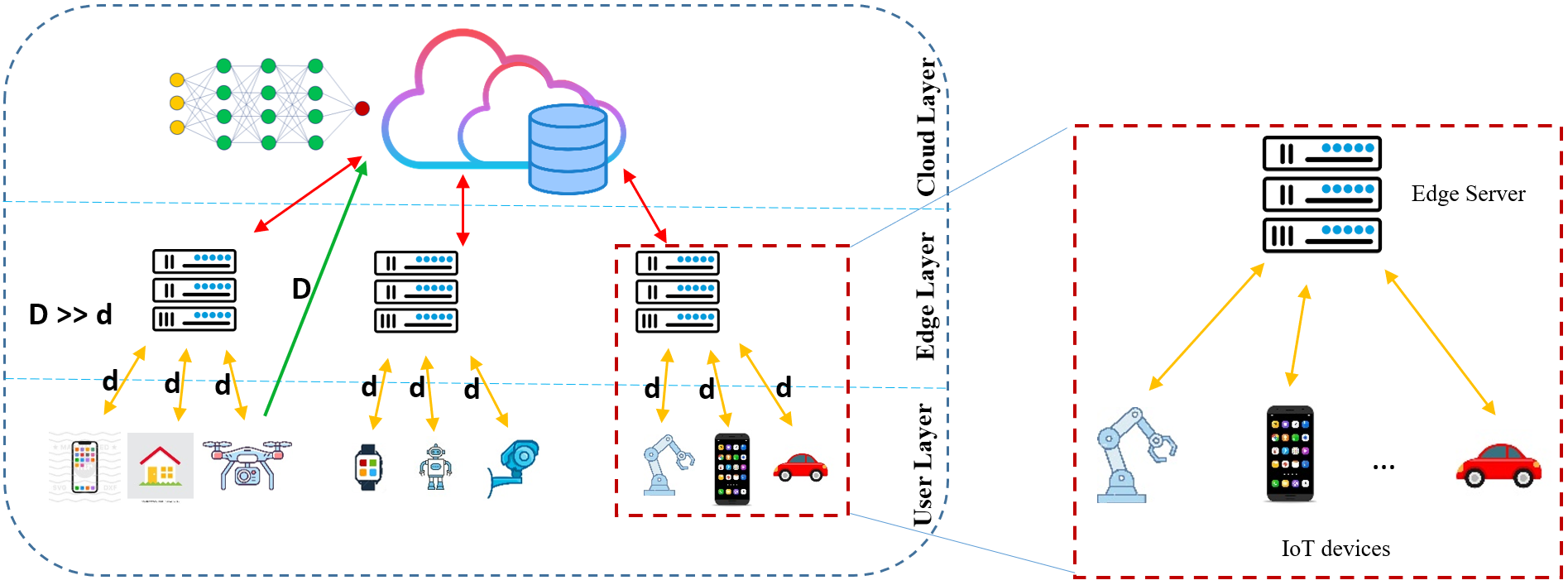}
\caption{System Architecture.}
\label{}
\end{figure*}

\textbf{FS phase:} The BQABC is a population-based method, and its procedure begins by initializing a random population. Each agent in this population is represented by a binary string, with a length exactly matching the number of original features in a dataset. The \(i\)th binary solution at the \(t\)th iteration of this algorithm can be represented as \(Z_i(t)=[z_i^1(t), z_i^2(t), ..., z_i^m(t)]\), where \(z_i^j(t)\in \{0,1\} \). In these strings, “1” indicates that the corresponding feature is selected, while “0” indicates it is not selected. Subsequently, these agents (primary solutions) are displaced, combined and evolved during the iterations. To achieve this, at each iteration, a learning algorithm’s feedback is utilized to evaluate and score the agents. In other words, the learning algorithm serves as a fitness function, with its accuracy considered as a fitness value for the agents. Ultimately, the best-performing agent is identified as the optimal feature subset.

\textbf{FC phase:} In this step, the GP algorithm is employed to generate a new informative feature. Unlike most other population-based methods, the GP algorithm represents individuals as trees, which serves as a robust representation for mathematical expressions. The process begins by initializing the population, wherein the selected features obtained from the previous phase \(Selected-F=\{f_1, f_2, ..., f_s\}\) (FS phase) are combined with mathematical operations such as \(+\), \(-\), \(\times\), \(sin\), and \(cos\). Each tree, or newly created feature, is then evaluated using a classifier, and the accuracy of that classifier is assigned to the corresponding constructed feature as a fitness value. The process continues until the maximum number of allowed generation is reached, during which trees are reproduced using crossover and mutation operators to update the population. Upon reaching the termination condition, the best individual is selected as the high-level constructed feature.

After these two steps, the best constructed feature from the FC step \(\{f_{c}\}\) is added to the best selected features that are obtained in the FS step (\(Selected-F=\{f_1, f_2, ..., f_s\}\)). This augmented feature set (\(Augmented-F=\{f_1, f_2, ..., f_s, f_{c}\}\)) is used for training a classifier to detect network attacks and calculate its accuracy as a measure of the IDS's performance. The pseudocode of the proposed method is given in Algorithm 2, and this process is illustrated in Fig. 2.

\begin{algorithm}
\caption{Pseudocode of the proposed method}\label{alg:cap}
\begin{algorithmic}[1]
\renewcommand{\algorithmicrequire}{\textbf{Input:}}
\renewcommand{\algorithmicensure}{\textbf{Output:}}
\Require The population size for BQABC (\(S\)), Number of original features (\(m\)), The population size for GP (\(S^\prime\)), The maximum number of iterations, Predefined mathematical operations
\Ensure The best agent representing the best selected features and the best program representing the best constructed feature
\newline
\State Initialize the population of BQABC randomly
\While{reaching the stopping criteria or maximum iteration}
\State Evaluate agents by a fitness function
\State Generate new quantum food sources and update agents' positions based on BQABC algorithm (Algorithm 1)
\EndWhile
\State Save the best agent representing the best selected features
\State Extract the best selected features from the best agent
\State Initialize the population of GP randomly by using the best selected features as variables and predefined mathematical operations
\While{reaching the stopping criteria or maximum iteration}
\State Evaluate programs/constructed features by a fitness function
\State Select one or two programs by a selection approach
\State Apply genetic operators including: reproduction, crossover, and mutation on the selected chromosomes
\EndWhile
\State Save the best program representing the best constructed feature
\State Add the best constructed feature to the best selected ones
\end{algorithmic}
\end{algorithm}

\begin{figure}
\includegraphics[width=\linewidth, trim={6cm 18cm 6cm 3cm},clip]{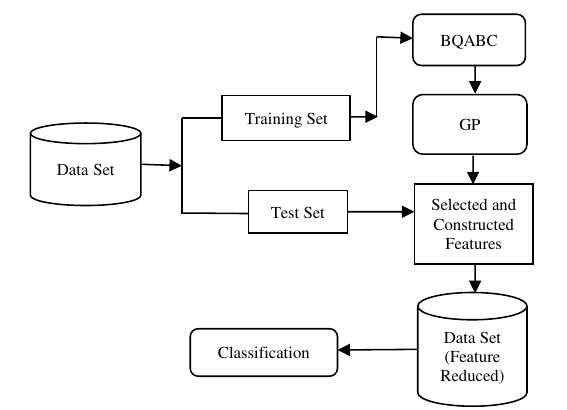}
\caption{Overview of the proposed method.}
\label{}
\end{figure}

\section{Experiments and Results}

\subsection{Dataset}
The proposed method is tested on three benchmark datasets: NSL-KDD \cite{tavallaee2009detailed}, UNSW-NB 15 \cite{moustafa2015unsw}, and Bot-IoT \cite{koroniotis2019towards}. The NSL-KDD dataset, an improved version of KDD99, comprises 41 features with 125,973 instances in the training set and 22,544 instances in the test set. The UNSW-NB 15 dataset includes 49 features created using the Argus tool. It is divided into a training set containing 175,341 samples and a testing set comprising 82,332 samples, covering various attack types and standard records. The Bot-IoT dataset is created to mimic a real-world scenario. This paper utilizes 5\(\%\) of the complete dataset, which comprises over 3.6 million instances \cite{kareem2022effective}. The characteristics of these datasets are indicated in Table I.

\begin{table}[htbp]
\caption{Characteristics of the datasets}

\begin{center}
\resizebox{\columnwidth}{!}{%
\begin{tabular}{l l l l}
\hline
Criteria ($\downarrow$)\(/\)Dataset name ($\rightarrow$) & NSL-KDD & UNSW-NB15 & BoT-IoT \\
\hline
Number of features & 41 & 49 & 43\\

Number of attack categories & 4 & 9 & 4\\

Number of classes & 5 & 10 & 5 \\

Number of total instances & 148517 & 257673 & 3668522 \\

Number of instances in training set & 125973 & 175341 & 2934817  \\

Number of instances in test set & 22544 & 82332 & 733705 \\
\hline
\end{tabular}
}
\label{tab1}
\end{center}
\end{table}

\subsection{Evaluation Measure}
In wrapper-based FS and FC methods, performance of a classifier is used to evaluate and score selected feature subsets or constructed features. In this paper, K-nearest neighbor (KNN) classifier is used as a fitness function in BQABC and GP algorithm. The performance metrics extracted from the confusion matrix and employed in this study are accuracy, sensitivity (recall), specificity, and the False Positive Rate (FPR). These evaluation metrics are calculated using equations (1) to (4).

\begin{equation}\label{my_eleven_eqn_first}
Accuracy = \frac{TP+TN}{TP+FP+FN+TN}
\end{equation}

\begin{equation}\label{my_eleven_eqn_second}
Sensitivity (recall) = \frac{TP}{TP+FN}
\end{equation}

\begin{equation}\label{my_eleven_eqn_second}
Specificity = \frac{TN}{TN+FP}
\end{equation}

\begin{equation}\label{my_eleven_eqn_second}
FPR = \frac{FP}{FP+TN}
\end{equation}

\noindent where \(TP\), \(TN\), \(FP,\) and \(FN\) denote true positive, true negative, false positive, and false negative, respectively.

\subsection{Parameter Setting}
The KNN Learning algorithm is employed as the classifier. Compared to more complex learning algorithms such as neural networks, KNN is straightforward yet performs well, often competing with other complex classifiers. Given the limitations of edge servers previously discussed, a simple classifier like KNN is used in this work. According to \cite{kareem2022effective}, the value of \(K\) in the KNN classifier, denoting the number of nearest neighbors, is set to 5 for result comparison. In BQABC, the minimum and maximum values of the rotation angle which is indicated by \(\theta\) are \(0.001\pi\) and \(0.05\pi\) respectively. The total number of agents in BQABC and GP is set to 50, and maximum number of iterations is 100. Each of these methods is executed 30 times, and the average evaluation values are computed based on these 30 runs. It should be mentioned that all these values are chosen based on trial and error. All these parameters are demonstrated in Table II. 

\begin{table}[htbp]
\caption{Parameters of the proposed method}

\begin{center}
\begin{tabular}{|c|c|}
\hline
\textbf{Parameter} & \textbf{\textit{Parameter value}} \\
\hline
\textbf{Population size of BQABC}& 50 \\
\hline
\textbf{Population size of GP}& 50 \\
\hline
\textbf{Maximum number of iterations}& 100 \\
\hline
\textbf{Number of runs}& 30 \\
\hline
\textbf{Number of neighbors in KNN}& \(K=5\) \\
\hline 
\textbf{Maximum value of the rotation angle}& $\theta=0.05\pi$ \\
\hline
\textbf{Minimum value of the rotation angle}& $\theta=0.001\pi$ \\
\hline
\textbf{Internal node’s functions}& \(+\), \(-\), \(\times\), \(sin\), \(cos\) \\
\hline 
\end{tabular}
\label{tab1}
\end{center}
\end{table}

\textbf{Data pre-processing:} Also, the data is initially pre-processed to ensure uniform formatting. This involves a two-step process: first, non-numeric values are converted to numeric values, referred to as numericalization. Following this, a normalization process is applied to bring all numerical columns onto a common scale range. Some variables, when measured on different scales, can introduce bias into the model. Given that we are working with NSL-KDD, UNSW-NB15, and BoT-IoT as benchmark datasets, it has been observed that each feature exhibits distinct ranges. Therefore, min-max normalization is chosen for consistency. This approach involves normalizing the features by subtracting the mean from each feature and then dividing it by its standard deviation. Min-max normalization scales the values to fall within the range of [0, 1].

\subsection{Results and Analysis}
In this section, we evaluate the effectiveness of the proposed hybrid feature selection and construction method in identifying the optimal feature subset and constructed feature by comparing it to various other metaheuristic feature selection algorithms, such as CS-PSO \cite{ghosh2019cs}, HHO \cite{heidari2019harris}, MVO \cite{mirjalili2016multi}, HGS \cite{yang2021hunger}, BSA \cite{meng2016new}, GTO \cite{abdollahzadeh2021artificial}, and GTO-BSA \cite{kareem2022effective} which are described below. The experiments make use of three intrusion detection datasets: NSL-KDD, UNSW-NB15, and BoT-IoT, representing a good combination of zero-day security threats.

\begin{itemize}
\item Cuckoo search-Particle swarm optimization (CS-PSO): It combines two algorithms, cuckoo search and particle swarm optimization, while integrating the concepts of local best and global best. Then informative features are chosen based on their fitness values to classify all the attacks accurately .
\item Harris Hawks Optimize (HHO): It is a nature-inspired algorithm that mimics the cooperative behavior and hunting style of Harris's hawks in the wild, known as the 'surprise pounce'. It is used to remove irrelevant features to improve the performance of IDSs.
\item Multi-Verse Optimizer (MVO): This algorithm is rooted in three cosmological principles: white holes, black holes, and wormholes. 
\item Hunger Games Search (HGS): It is crafted to align with the hunger-driven behaviors and choices observed in animals.
\item Bird Swarm Algorithm (BSA): It is derived from swarm intelligence observed in bird swarms, encompassing three primary behaviors in birds: foraging, vigilance, and flight.
\item Gorilla Troops Optimizer (GTO): This algorithm mathematically models the collective life of gorillas and introduces novel mechanisms for both exploration and exploitation.
\item  Gorilla Troops Optimizer based on Bird Swarm Algorithm (GTO-BSA): This algorithm utilizes BSA to enhance the exploitation performance of GTO due to its strong capability in locating feasible regions with optimal solutions. Consequently, this contributes to an improved final output quality and enhances the performance of IDSs.
\end{itemize}

\textbf{Comparison with state of the art methods:} Tables III, IV, and V present the experimental results for NSL-KDD, UNSW-NB15, and BoT-IoT, respectively. The results indicate that, in nearly all evaluation metrics, the proposed hybrid feature selection and feature construction technique outperforms recent wrapper-based feature selection techniques with KNN-based IDS for all three intrusion detection datasets. In the case of the NSL-KDD dataset, the proposed approach attained an accuracy of 0.9889 with a reduced dataset containing 11 features. This reflects an approximate increase in accuracy of 0.033 to 0.0449 with the proposed hybrid method when applied to a KNN-based IDS using the NSL-KDD dataset. Meanwhile, on the UNSW-NB15 dataset, the proposed approach achieved an accuracy of 0.9022 while reducing the feature set by approximately 78.36\(\%\). As a result, an estimated increase in accuracy of approximately 0.1921 to 0.2696 is observed with the proposed method. Moving on to the BoT-IoT dataset, the proposed method achieved an accuracy of 0.9849 using a reduced feature subset consisting of about 10 features out of the original 43. This resulted in an estimated accuracy increase of around 0.0317 to 0.0525. However, the proposed method selects and constructs more features for this dataset compared to the other FS methods being compared.

A high False Positive Rate (FPR) can trigger false alarms, potentially causing the intrusion detection system to misclassify normal network traffic as malicious. With the proposed approach, it achieves promising scores for sensitivity (True Positive Rate), specificity (True Negative Rate), and FPR evaluation metrics across all three datasets. This method concurrently enhances both the True Positive Rate and True Negative Rate while reducing the FPR. Compared to other methods, the NSL-KDD dataset exhibits an increase in sensitivity of approximately 0.0561 to 0.0809 and an increase in specificity of around 0.0121 to 0.0169. These improvements result in a sensitivity of 0.9703 and a specificity of 0.9876 when applying the proposed hybrid technique. Similarly, for the UNSW-NB15 dataset, there is an increase in sensitivity ranging from about 0.133 to 0.2656 and an improvement in specificity ranging from approximately 0.0023 to 0.0779, culminating in a sensitivity of 0.9483 and a specificity of 0.8806. Nonetheless, in the case of the BoT-IoT dataset, the HHO method achieved a higher sensitivity of 0.9992, while the sensitivity of the proposed method was 0.9979. On the other hand, the proposed method obtained the highest specificity of 0.9927, whereas HHO's specificity was 0.5111. Furthermore, the proposed method has demonstrated superior performance in terms of achieving the lowest FPR across all three datasets.

\textbf{Computational time analysis:} Table VI illustrates the average computational time (\(s\)) across 30 repetitions for the compared methods. Given our focus on the performance-cost trade-off, it is essential to simultaneously consider Table VI alongside Tables III, IV, and V for a comprehensive understanding. By comparing table VI and table III, it is clear that the proposed method outperforms the overall best-performing algorithm (GTO-BSA) in terms of speed by a factor of 7 while also enhancing accuracy from 0.9559 to 0.9889. This comparison highlights that the proposed method offers a balanced trade-off between computational cost and other performance metrics for almost all datasets. For example, by considering Table VI and Table V, it becomes apparent that the proposed method outperforms the HHO method in computational cost by a factor of 1.5. Furthermore, the sensitivity remains nearly identical (0.9979), while other performance metrics exhibit improvement.

\begin{table}[htbp]
\caption{Comparison of the proposed method with existing methods in the literature on the NSL-KDD dataset.}

\begin{center}
\resizebox{\columnwidth}{!}{%
\begin{tabular}{l l c c c c}
\hline
Method & (\#) Features & Accuracy & Sensitivity & Specificity & FPR \\ \hline
CS-PSO & 14.875 & 0.9501 & 0.8894 & 0.9755 & 0.0245 \\ \hline
HHO & 19.625 & 0.9544 & 0.9059 & 0.9733 & 0.0267 \\ \hline
MVO & 16.5 & 0.9531 & 0.8981 & 0.9749 & 0.0251 \\ \hline
HGS & 18.625 & 0.9479 & 0.8945 & 0.9743 & 0.0257 \\ \hline
BSA & 21.125 & 0.9440 & 0.9003 & 0.9707 & 0.0293 \\ \hline
GTO & 18.5 & 0.9543 & 0.9032 & 0.9738 & 0.0262 \\ \hline
GTO-BSA & 14.75 & 0.9559 & 0.9142 & 0.9736 & 0.0264 \\ \hline
Proposed method & \textbf{11} & \textbf{0.9889} & \textbf{0.9703} & \textbf{0.9876} & \textbf{0.0124} \\ \hline

\end{tabular}
}
\label{tab1}
\end{center}

\end{table}

\begin{table}[htbp]
\caption{Comparison of the proposed method with existing methods in the literature on the UNSW-NB15 dataset.}

\begin{center}
\resizebox{\columnwidth}{!}{%
\begin{tabular}{l l c c c c}
\hline
Method & (\#) Features & Accuracy & Sensitivity & Specificity & FPR \\ \hline
CS-PSO & 18.125 & 0.6692 & 0.7730 & 0.8663 & 0.1337 \\ \hline
HHO & 12 & 0.7064 & 0.7865 & 0.8222 & 0.1778 \\ \hline
MVO & 16.5 & 0.6979 & 0.7807 & 0.8783 & 0.1217 \\ \hline
HGS & 18.75 & 0.6326 & 0.6827 & 0.8401 & 0.1599 \\ \hline
BSA & 14.875 & 0.6543 & 0.7519 & 0.8675 & 0.1325 \\ \hline
GTO & 12.625 & 0.7072 & 0.7788 & 0.8027 & 0.1973 \\ \hline
GTO-BSA & 16.625 & 0.7101 & 0.8153 & 0.8770 & 0.1230 \\ \hline
Proposed method & \textbf{10.6} & \textbf{0.9022} & \textbf{0.9483} & \textbf{0.8806} & \textbf{0.1194} \\ \hline
\end{tabular}
}
\label{tab1}
\end{center}

\end{table}

\begin{table}[htbp]
\caption{Comparison of the proposed method with existing methods in the literature on the BoT-IoT dataset.}

\begin{center}
\resizebox{\columnwidth}{!}{%
\begin{tabular}{l l c c c c}
\hline
Method & (\#) Features & Accuracy & Sensitivity & Specificity & FPR \\ \hline
CS-PSO & 3.2 & 0.9370 & 0.9648 & 0.9284 & 0.0716 \\ \hline
HHO & \textbf{2.133} & 0.9532 & \textbf{0.9992} & 0.5111 & 0.4889 \\ \hline
MVO & 2.8 & 0.9393 & 0.9670 & 0.8584 & 0.1416 \\ \hline
HGS & 3.6 & 0.9351 & 0.9620 & 0.9277 & 0.0723 \\ \hline
BSA & 3.4 & 0.9324 & 0.9512 & 0.6504 & 0.3496 \\ \hline
GTO & 2.4666 & 0.9479 & 0.9885 & 0.6500 & 0.3500 \\ \hline
GTO-BSA & 2.5333 & 0.9485 & 0.9928 & 0.9622 & 0.0378 \\ \hline
Proposed method & 10.33 & \textbf{0.9849} & 0.9979 & \textbf{0.9927} & \textbf{0.0073} \\ \hline
\end{tabular}
}
\label{tab1}
\end{center}

\end{table}

\begin{table}[htbp]
\caption{Comparison of the proposed method with existing methods in the literature in terms of computational time (\(s\))}

\begin{center}
\begin{tabular}{l c c c}
\hline
Method ($\downarrow$)\(/\)Dataset name ($\rightarrow$)  & NSL-KDD & UNSW-NB15 & BoT-IoT \\ \hline
CS-PSO & 4604.31 & 80.89 & 71.09\\ \hline
HHO & 12,476.16 & 146.24 & 144.71\\ \hline
MVO & 5441.159 & 77.87 & 70.17\\ \hline
HGS & \textbf{661.72} & \textbf{12.57} & \textbf{10.74}\\ \hline
BSA & 6515.84 & 74.88 & 68.97 \\ \hline
GTO & 9719.66 & 113.36 & 108.63 \\ \hline
GTO-BSA & 10,205.83 & 161.23 & 145.74 \\ \hline
Proposed method & 1441.02 & 133.55 & 92.50 \\ \hline
\end{tabular}
\label{tab1}
\end{center}

\end{table}

\begin{table}[t]
\caption{Results of the Wilcoxon signed-rank test for the proposed method}

\begin{center}
\begin{tabular}{l c }
\hline
Proposed method VS. & Exact P-value  \\ \hline
CS-PSO & 0.003906 \\ 
HHO & 0.007812 \\ 
MVO & 0.003906 \\ 
HGS & 0.003906 \\ 
BSA & 0.003906 \\ 
GTO & 0.003906 \\ 
GTO-BSA & 0.003906 \\ \hline
\end{tabular}
\label{tab1}
\end{center}

\end{table}

\textbf{Non-parametric statistical analysis:} The proposed approach enhances IDS performance in most cases by carefully choosing and creating informative features from the original feature set. To statistically validate the achieved results, the nonparametric Wilcoxon signed-rank test is utilized for all the performance measures. In Table VII, the exact p-values are presented, confirming significant differences between the proposed hybrid method and the other compared algorithms. According to this test, any methods with p-values less than 0.05 are rejected. Notably, the proposed method rejects all of the comparative algorithms.

\section{Conclusion and Future Works}
In this paper, we introduce a novel feature engineering method to find a balance trade-off between cost and accuracy to enhance intrusion detection system (IDS) performance. This approach focuses on increasing detection accuracy while concurrently reducing false positive rates by identifying the most informative features that positively impact IDS performance. We evaluate the performance of the proposed feature engineering method using three IoT intrusion detection datasets: NSL-KDD, UNSW-NB15, and BoT-IoT, and compare it with other competitive algorithms. The results indicate achieved accuracies of 0.9889, 0.9022, and 0.9849 for the NSL-KDD, UNSW-NB15, and BoT-IoT datasets, respectively. As this study marks the initial exploration of feature construction in the context of intrusion detection, future research endeavors may extend this approach to construct multiple features, potentially replacing the total number of original features with the newly created ones. Furthermore, in future endeavors, our focus may shift towards detecting attacks within a distributed IoT environment, rather than concentrating solely on a single edge server.
\section*{Acknowledgement}
This work is funded by research grant provided by the National Science Foundation (NSF) under the grant number 1948387.

\bibliographystyle{IEEEtranN}
\bibliography{References}

\begin{thebibliography}{21}
\providecommand{\natexlab}[1]{#1}
\providecommand{\url}[1]{#1}
\csname url@samestyle\endcsname
\providecommand{\newblock}{\relax}
\providecommand{\bibinfo}[2]{#2}
\providecommand{\BIBentrySTDinterwordspacing}{\spaceskip=0pt\relax}
\providecommand{\BIBentryALTinterwordstretchfactor}{4}
\providecommand{\BIBentryALTinterwordspacing}{\spaceskip=\fontdimen2\font plus
\BIBentryALTinterwordstretchfactor\fontdimen3\font minus \fontdimen4\font\relax}
\providecommand{\BIBforeignlanguage}[2]{{%
\expandafter\ifx\csname l@#1\endcsname\relax
\typeout{** WARNING: IEEEtranN.bst: No hyphenation pattern has been}%
\typeout{** loaded for the language `#1'. Using the pattern for}%
\typeout{** the default language instead.}%
\else
\language=\csname l@#1\endcsname
\fi
#2}}
\providecommand{\BIBdecl}{\relax}
\BIBdecl

\bibitem[Albulayhi et~al.(2022)Albulayhi, Abu Al-Haija, Alsuhibany, Jillepalli, Ashrafuzzaman, and Sheldon]{albulayhi2022iot}
K.~Albulayhi, Q.~Abu Al-Haija, S.~A. Alsuhibany, A.~A. Jillepalli, M.~Ashrafuzzaman, and F.~T. Sheldon, ``Iot intrusion detection using machine learning with a novel high performing feature selection method,'' \emph{Applied Sciences}, vol.~12, no.~10, p. 5015, 2022.

\bibitem[Mahanipour and Khamfroush(2023)]{mahanipour2023wrapper}
A.~Mahanipour and H.~Khamfroush, ``Wrapper-based federated feature selection for iot environments,'' in \emph{2023 International Conference on Computing, Networking and Communications (ICNC)}.\hskip 1em plus 0.5em minus 0.4em\relax IEEE, 2023, pp. 214--219.

\bibitem[Mishra and Paliwal(2023)]{mishra2023mitigating}
A.~K. Mishra and S.~Paliwal, ``Mitigating cyber threats through integration of feature selection and stacking ensemble learning: the lgbm and random forest intrusion detection perspective,'' \emph{Cluster Computing}, vol.~26, no.~4, pp. 2339--2350, 2023.

\bibitem[Al-Yaseen et~al.(2022)Al-Yaseen, Idrees, and Almasoudy]{al2022wrapper}
W.~L. Al-Yaseen, A.~K. Idrees, and F.~H. Almasoudy, ``Wrapper feature selection method based differential evolution and extreme learning machine for intrusion detection system,'' \emph{Pattern Recognition}, vol. 132, p. 108912, 2022.

\bibitem[Kareem et~al.(2022)Kareem, Mostafa, Hashim, and El-Bakry]{kareem2022effective}
S.~S. Kareem, R.~R. Mostafa, F.~A. Hashim, and H.~M. El-Bakry, ``An effective feature selection model using hybrid metaheuristic algorithms for iot intrusion detection,'' \emph{Sensors}, vol.~22, no.~4, p. 1396, 2022.

\bibitem[Barani and Nezamabadi-pour(2018)]{barani2018bqiabc}
F.~Barani and H.~Nezamabadi-pour, ``Bqiabc: a new quantum-inspired artificial bee colony algorithm for binary optimization problems,'' \emph{Journal of AI and Data Mining}, vol.~6, no.~1, pp. 133--143, 2018.

\bibitem[Koza(1994)]{koza1994genetic}
J.~R. Koza, ``Genetic programming as a means for programming computers by natural selection,'' \emph{Statistics and computing}, vol.~4, pp. 87--112, 1994.

\bibitem[Mahanipour and Nezamabadi-Pour(2019)]{mahanipour2019multiple}
A.~Mahanipour and H.~Nezamabadi-Pour, ``A multiple feature construction method based on gravitational search algorithm,'' \emph{Expert Systems with Applications}, vol. 127, pp. 199--209, 2019.

\bibitem[Thakkar and Lohiya(2023)]{thakkar2023fusion}
A.~Thakkar and R.~Lohiya, ``Fusion of statistical importance for feature selection in deep neural network-based intrusion detection system,'' \emph{Information Fusion}, vol.~90, pp. 353--363, 2023.

\bibitem[Kumar et~al.(2020)Kumar, Sinha, Das, Pandey, and Goswami]{kumar2020integrated}
V.~Kumar, D.~Sinha, A.~K. Das, S.~C. Pandey, and R.~T. Goswami, ``An integrated rule based intrusion detection system: analysis on unsw-nb15 data set and the real time online dataset,'' \emph{Cluster Computing}, vol.~23, pp. 1397--1418, 2020.

\bibitem[Liu and Shi(2022)]{liu2022hybrid}
Z.~Liu and Y.~Shi, ``A hybrid ids using ga-based feature selection method and random forest,'' \emph{Int. J. Mach. Learn. Comput}, vol.~12, no.~02, pp. 43--50, 2022.

\bibitem[Nazir and Khan(2021)]{nazir2021novel}
A.~Nazir and R.~A. Khan, ``A novel combinatorial optimization based feature selection method for network intrusion detection,'' \emph{Computers \& Security}, vol. 102, p. 102164, 2021.

\bibitem[Tavallaee et~al.(2009)Tavallaee, Bagheri, Lu, and Ghorbani]{tavallaee2009detailed}
M.~Tavallaee, E.~Bagheri, W.~Lu, and A.~A. Ghorbani, ``A detailed analysis of the kdd cup 99 data set,'' in \emph{2009 IEEE symposium on computational intelligence for security and defense applications}.\hskip 1em plus 0.5em minus 0.4em\relax Ieee, 2009, pp. 1--6.

\bibitem[Moustafa and Slay(2015)]{moustafa2015unsw}
N.~Moustafa and J.~Slay, ``Unsw-nb15: a comprehensive data set for network intrusion detection systems (unsw-nb15 network data set),'' in \emph{2015 military communications and information systems conference (MilCIS)}.\hskip 1em plus 0.5em minus 0.4em\relax IEEE, 2015, pp. 1--6.

\bibitem[Koroniotis et~al.(2019)Koroniotis, Moustafa, Sitnikova, and Turnbull]{koroniotis2019towards}
N.~Koroniotis, N.~Moustafa, E.~Sitnikova, and B.~Turnbull, ``Towards the development of realistic botnet dataset in the internet of things for network forensic analytics: Bot-iot dataset,'' \emph{Future Generation Computer Systems}, vol. 100, pp. 779--796, 2019.

\bibitem[Ghosh et~al.(2019)Ghosh, Karmakar, Sharma, and Phadikar]{ghosh2019cs}
P.~Ghosh, A.~Karmakar, J.~Sharma, and S.~Phadikar, ``Cs-pso based intrusion detection system in cloud environment,'' in \emph{Emerging Technologies in Data Mining and Information Security: Proceedings of IEMIS 2018, Volume 1}.\hskip 1em plus 0.5em minus 0.4em\relax Springer, 2019, pp. 261--269.

\bibitem[Heidari et~al.(2019)Heidari, Mirjalili, Faris, Aljarah, Mafarja, and Chen]{heidari2019harris}
A.~A. Heidari, S.~Mirjalili, H.~Faris, I.~Aljarah, M.~Mafarja, and H.~Chen, ``Harris hawks optimization: Algorithm and applications,'' \emph{Future generation computer systems}, vol.~97, pp. 849--872, 2019.

\bibitem[Mirjalili et~al.(2016)Mirjalili, Mirjalili, and Hatamlou]{mirjalili2016multi}
S.~Mirjalili, S.~M. Mirjalili, and A.~Hatamlou, ``Multi-verse optimizer: a nature-inspired algorithm for global optimization,'' \emph{Neural Computing and Applications}, vol.~27, pp. 495--513, 2016.

\bibitem[Yang et~al.(2021)Yang, Chen, Heidari, and Gandomi]{yang2021hunger}
Y.~Yang, H.~Chen, A.~A. Heidari, and A.~H. Gandomi, ``Hunger games search: Visions, conception, implementation, deep analysis, perspectives, and towards performance shifts,'' \emph{Expert Systems with Applications}, vol. 177, p. 114864, 2021.

\bibitem[Meng et~al.(2016)Meng, Gao, Lu, Liu, and Zhang]{meng2016new}
X.-B. Meng, X.~Z. Gao, L.~Lu, Y.~Liu, and H.~Zhang, ``A new bio-inspired optimisation algorithm: Bird swarm algorithm,'' \emph{Journal of Experimental \& Theoretical Artificial Intelligence}, vol.~28, no.~4, pp. 673--687, 2016.

\bibitem[Abdollahzadeh et~al.(2021)Abdollahzadeh, Soleimanian~Gharehchopogh, and Mirjalili]{abdollahzadeh2021artificial}
B.~Abdollahzadeh, F.~Soleimanian~Gharehchopogh, and S.~Mirjalili, ``Artificial gorilla troops optimizer: a new nature-inspired metaheuristic algorithm for global optimization problems,'' \emph{International Journal of Intelligent Systems}, vol.~36, no.~10, pp. 5887--5958, 2021.

\end{thebibliography}

\end{document}